# Non-English language publications in Citation Indexes – quantity and quality.

Olga Moskaleva[1] and Mark Akoev [2]

[1] *o.moskaleva@spbu.ru*
Saint-Petersburg State University, Universitetskaya emb. 7-9, 199034, St. Petersburg, Russia

[2] *m.a.akoev@urfu.ru*
Ural Federal University, 19 Mira str., Ekaterinburg, 620002, Russia

**Abstract.**

We analyzed publications data in WoS and Scopus to compare publications in native languages vs publications in English and find any distinctive patterns. We analyzed their distribution by research areas, languages, type of access and citation patterns. The following trends were found: share of English publications increases over time; native-language publications are read and cited less than English-language outside the origin country; open access impact on views and citation is higher for native languages; journal ranking correlates with the share of English publications for multi-language journals. We conclude also that the role of non-English publications in research evaluation in non-English speaking countries is underestimated when research in social science and humanities is assessed only by publications in Web of Science and Scopus.

**Introduction**

The concept of English language as the modern lingua franca in science (Montgomery, 2016) is suggested by many studies. The number and proportion of English-language documents in Web of Science is growing (Kumar, Panwar, & Mahesh, 2016; W. Liu, 2017; Meneghini & Packer, 2007). Nevertheless, the total number of non-English publications is growing also as the number of researchers in non-English speaking countries (UNESCO, 2016) grows.

Available publications we have found focus mainly on the number of non-English language documents or the disadvantage which they receive in the number of citations (Di Bitetti & Ferreras, 2017; Fung, 2008) or journal impact factors (Diekhoff, Schlattmann, & Dewey, 2013; González-Alcaide, Valderrama-Zurián, & Aleixandre-Benavent, 2012; Liang, Rousseau, & Zhong, 2013; F. Liu, Hu, Tang, & Liu, 2018). Presumably, journal articles were analyzed.

The dominance of English language appears to be crucial at the second half of XX century, though in 1960-1970 Russian Language was the second in Natural Sciences (Ammon, 2012) after English, while at the beginning of XX English, French and German were almost in equal proportions. In Social Sciences, according to International Bibliography of the Social Sciences, French is the most popular scientific language together with English (Ammon, 2012). The same article underlines preferences of English language regardless of scientist's native languages – publications in English in almost all cases have a higher impact.

Michael D.Gordin (Gordin, 2015) gives a brilliant review of the role of different languages in science in a monograph. He describes reasons and consequences of English language dominance in scientific literature, differences in natural and social sciences, etc.

There are more than seven thousands of languages in the world (though not all of them have a written form). The most widely spread language by the number of speakers is Chinese (1,2 billion), the second is Spanish, the third – English, but the number of countries where English is the main language is three times higher than Spanish – 106 English-speaking countries against 31 for Spanish (Ethnologue, 2015).

**Non-English language publications in citation indexes**

The number of non-English documents in Web of Science or Scopus does not correlate neither with number of speakers, nor with the number of countries. We can see no correlation with the number of researchers with corresponding native language either. In 2013 there were 7,8 million researchers, among them 25% with English being native or one of state languages, Chinese - 17%, Japanese - 8,5%, Russian - 6%, German - 5%, Spanish - 3% and Portugal - 2% (UNESCO, 2016).

On the other hand, many research publications are not indexed in global citation databases, such as Web of Science and Scopus, especially in Social Science and Humanities (Kulczycki et al., 2018). The study of publication output in these areas in 7 European non-English speaking countries demonstrated that only part of them are covered by Web of Science (50,4% for Denmark and 15% for Poland), so research evaluation may be not correct when using only Web of Science data. This obstacle is the serious reason for national citation databases emerging globally (Pislyakov, 2007), enabling more adequate research assessment, especially in Social science & Humanities (Sīle, Guns, Sivertsen, & Engels, 2017). Some of independently created national databases become basis for national citation indexes on Web of Science platform, i.e. Chinese Citation Index, SciELO, Korean Citation Index and Russian Science Citation Index (Moskaleva, Pislyakov, Sterligov, Akoev, & Shabanova, 2018). By the end of May 2018 Clarivate Analytics announced plans for creating of a new citation index: Arabic Citation Index (ARCI) is to be launched in 2020 ("Clarivate Analytics Partners with the Egyptian Knowledge Bank to Power the First Arabic Citation Index - Clarivate," 2018).

In contrast to the opinion widely spread in non-English countries, global citation indexes include documents not only in English, and thanks to the document indexing policies, only the best material (journals, conference proceedings, books, etc.) is presented in Web of Science and Scopus. This obstacle makes it possible to assess the national science in total and the role of national languages in global science. A lot of information about publications in national languages can be extracted from reference lists in indexed documents, so everybody can draw their own opinion about national publications, cited in thoroughly selected sources.

The comparison of language-speaking countries and countries with publications in certain languages demonstrates more language diversity for research communication (Fig.1). Certainly, mainly it happens due to international collaboration, the main language of which is English (not shown on diagram).

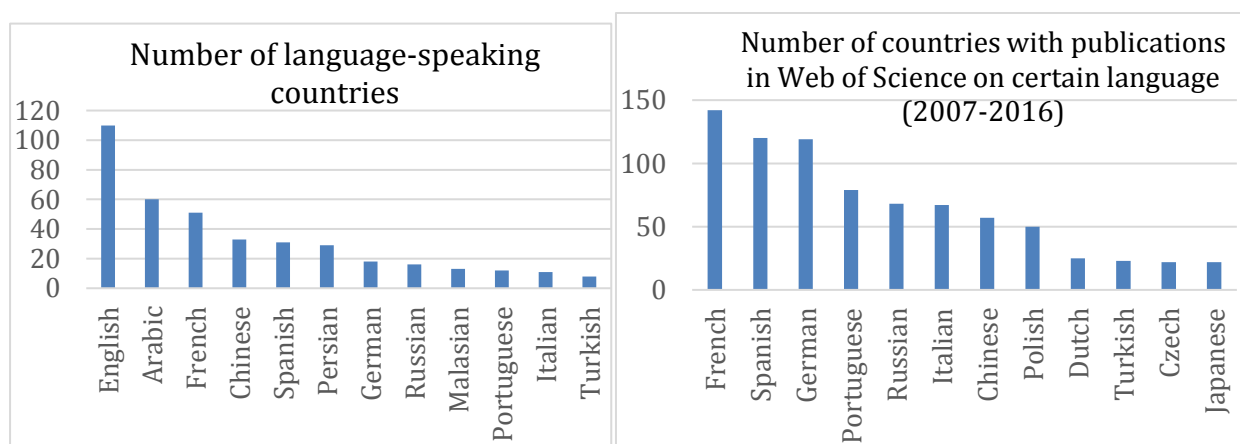

**Fig.1. Ranking of language-speaking countries and affiliated countries in Web of Science publications.**

The total number of non-English publications is growing from the beginning of XXI century both in Web of Science CC and in Scopus, though their share in total output diminishes (Fig.2). The proportion of English and non-English publications differs in various indexes in Web of Science CC and different publication types in Scopus. Less representation of non-English language journals and, as a consequence, of non-Anglophone publications in Web of Science CC as to compare with Scopus is compensated by development of national citation indexes on Web of Science platform. Such indexes are fully integrated with all other databases on Web of Science platform (Core Collection, BIOSIS, Medline, Derwent, etc.), so researchers in the whole world can have access to non-English language science of a high quality. Using Web of Science or Scopus as a search engine for publications provides researchers content of a higher quality in comparison with Google Scholar, for example, because of thoroughly selected sources and vast range of refine possibilities. All these tools help researcher to find relevant content with less efforts and time independently of original language of publication.

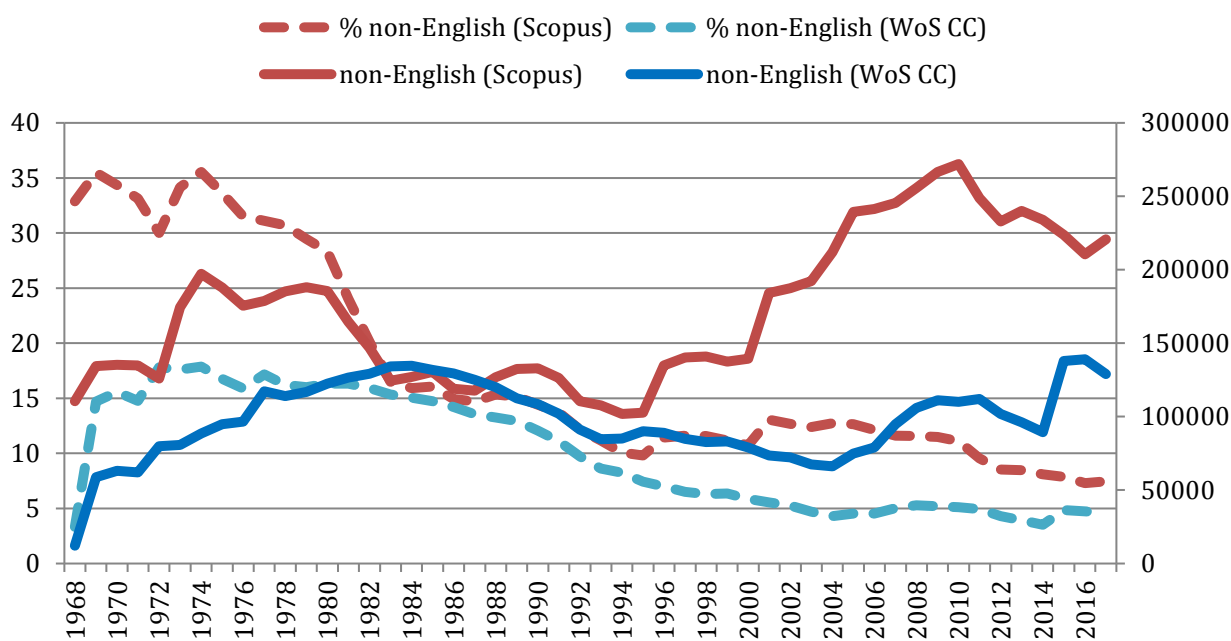

**Fig.2 The total number and proportion of non-English documents in Web of Science CC and Scopus**

Thus the international language in science now remains English, as earlier it was Latin, French or German. The growing publication output and proportion of English-language documents in global citation databases approves this fact. The common language for science makes an easier communication between researchers from different countries possible due to the common terminology. It allows attracting more scientists to the development of scientific knowledge, and on the other hand allows scientists with fluent English to have access to all scientific knowledge, and not just to a subset that is translated into national languages.

There are significant differences between research areas with respect to the role of non-English language documents. In Web of Science Core Collection for example, the share of non-English language publications in Arts & Humanities Citation Index exceeds 25% of total research output. Relatively high proportion of non-English documents is a characteristic feature of Emerging Sources Citation Index. Conference Proceedings in Social Science & Humanities (CPCI-SSH) also demonstrate 7 times higher proportion of non-English documents than in Natural Sciences and Engineering (CPCI-S). The same holds true for Book Citation Indexes (Fig.3). The dynamics of non-English documents in citation indexes in Web of Science CC also differs significantly (W. Liu, 2017).

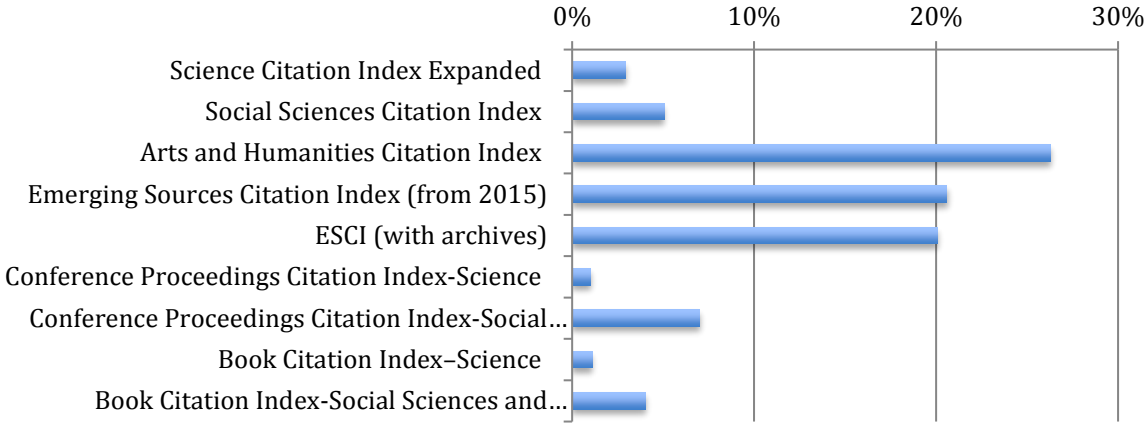

**Fig.3 Distribution of non-English language documents among citation indexes in Web of Science CC (2007-2016, % of total number of documents ).**

In Scopus we can see differences in dynamic of the Non-English publication share for different source types. It is decreasing for journals and relatively stable for books and conference proceedings (Fig.4).

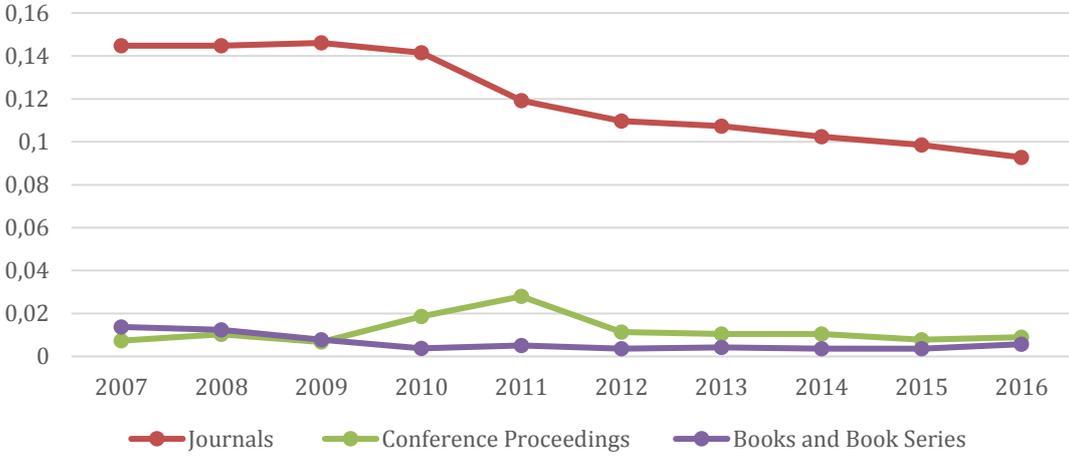

**Fig.4 The proportion of non-English language documents in different source types in Scopus**

The distribution of non-English documents by languages in Web of Science and Scopus looks very similar except for Chinese leading in Scopus and being only on the fourth place in Web of Science (Fig.5).

Even here we can see differences among different citation indexes in Web of Science Core Collection – German dominates in SCI-E, SSCI, BKCI-S and BKCI-SSH, the main language after English in AHCI is French, in CPCI-S and CPCI-SSH the main language is Chinese. In ESCI the highest share of non-English documents belongs to documents in Spanish.

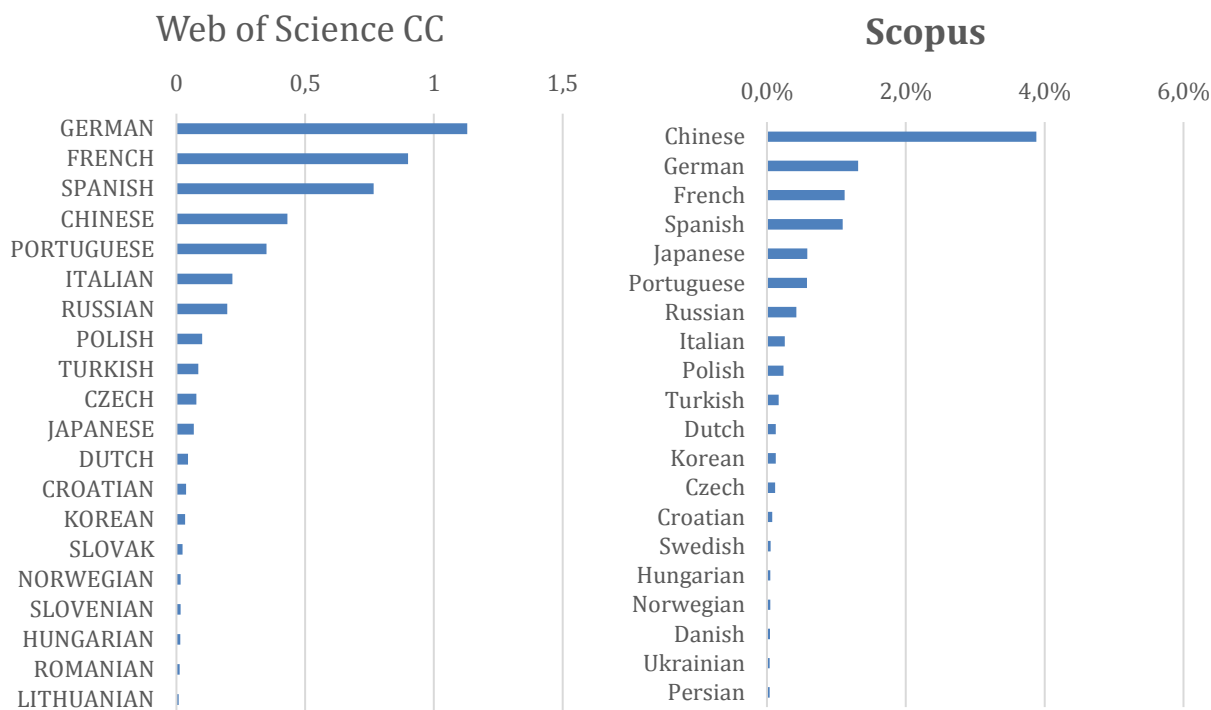

**Fig.5 The language distribution of non-English language documents in Web of Science CC and Scopus (2007-2016, % of total number of documents).**

Large body of research demonstrates the growing role of open access, which now influences both visibility and citedness of scientific results. New options in citation indexes were launched in 2017 – beginning of 2018 that allow finding articles available in gold (both in Web of Science and in Scopus) and greening (only in Web of Science) open access. These options make it easier to analyze their role in science development. The "elder" journal indexes of Web of Science CC (SCI-E, SSCI, AHCI) include up to 25% of open access documents (in SCI-E), less in CPCI and BKCI. Among non-English language publications, the best proportion of OA is observed for SCI-E – over 15%. ESCI is the most "open" citation database in Web of Science CC and includes 35% of OA documents, with more than 27% of non-English language publications available in open access (Fig.6).

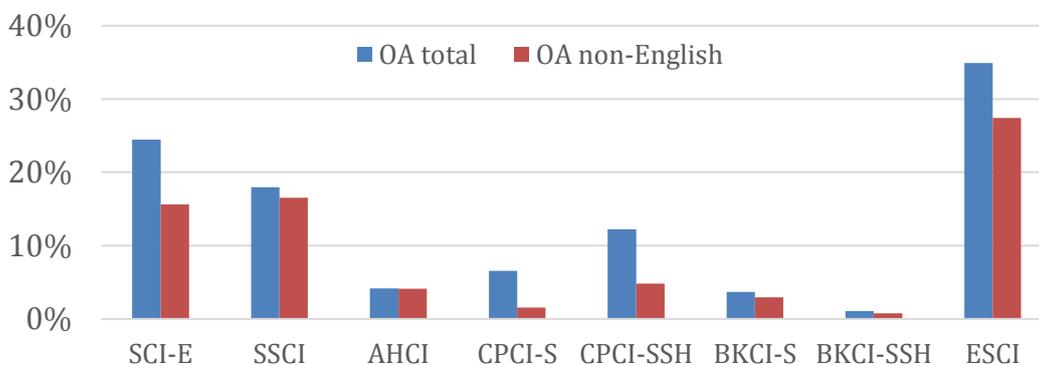

**Fig.6 The proportion of Gold open access publications in Web of Science CC (2007-2016)**

The languages distribution of OA publications by the share of OA documents in the total output in the same language is demonstrated on Fig.7.

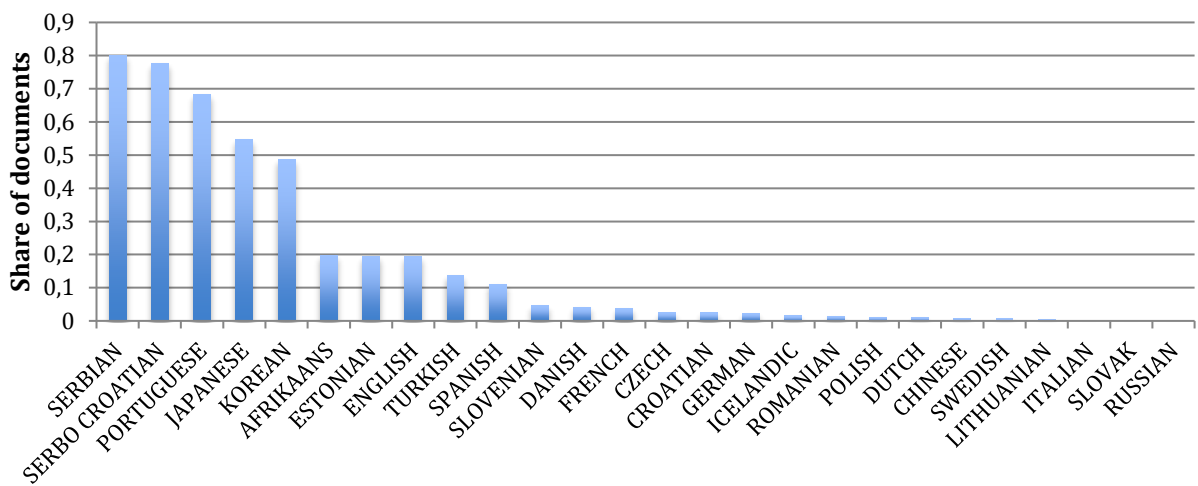

**Fig.7 The proportion of Gold OA in journal papers in different languages (2006-2017, SCI-E, SSCI, AHCI)**

Serbian, Portuguese, Japanese and Korean languages appeared to be the most open among other non-English papers in journals of "elder" journal indexes of Web of Science CC with 50-80% of Gold OA documents.

To determine the citation impact of open access publications we created corresponding data sets and analyzed them in InCites – analytical instrument of Clarivate Analytics. It provides a lot of complex bibliometric indicators and allows comparing different publication datasets independently of publication date, source and subject category. The results on Fig.8 show that both average citation and CNCI are almost twice higher for OA publications in Portuguese and Japanese than for traditional publications that are available only with subscription. At the same time, there are almost no non-English language publications among highly cited documents, independently of access type. In total, their citation is much lower than of English-language documents.

Researchers from different countries use their national languages to present their results to a different extent. The proportion of documents in Portuguese in Brazil is almost 14%, while in Portugal only 3% of total number of documents in Web of Science CC are in Portuguese (Fig.9).

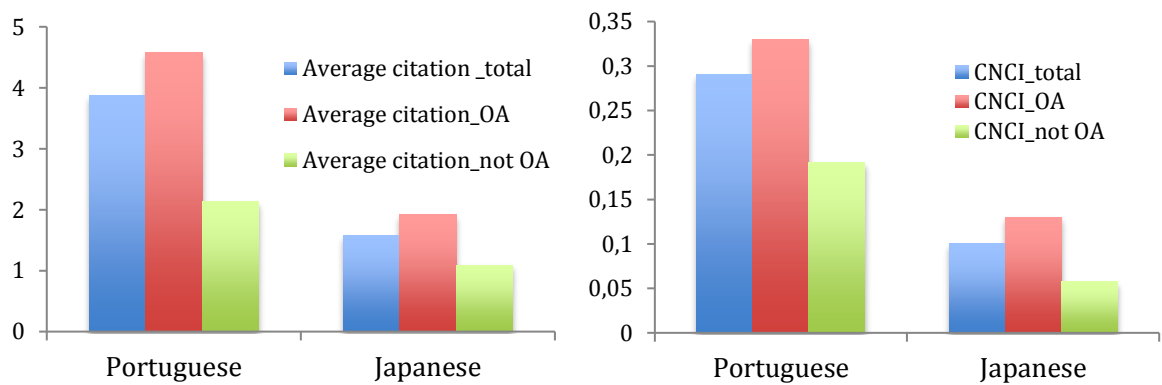

**Fig.8 Average citation and Category Normalized citation impact (CNCI) of journal articles in Japanese and Portuguese languages with different access types (Web of Science, 2008-2016, SCI-E, SSCI, AHCI)**

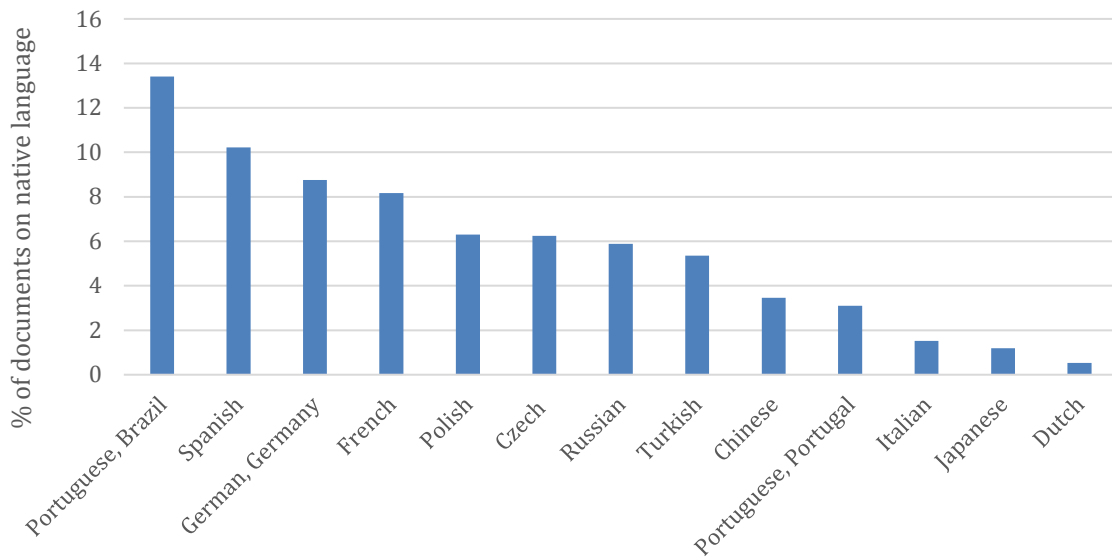

**Fig.9 The proportion of documents in native languages for different countries (Web of Science CC, 2006-2017)**

Native-language publications are cited less than English-language almost in all examined cases (Fig.10), but in Humanities this influence is lower than for STM. In Brazil, for example, the CNCI of Portuguese publications in Humanities is the same or for some years even higher than for total country documents in this area. CNCI of Spanish-language documents in Humanities is 1,76, with 1,54 for all country publications in Humanities. Influence of language on citation rate is described in literature, though the investigations are not numerous (Diekhoff et al., 2013; Fung, 2008; Shu & Larivière, 2015). Together with data on large proportion of non-English language publications in SSCI and AHCI (Fig.3) it can demonstrate the great role of native-language publications in Social Sciences and Humanities.

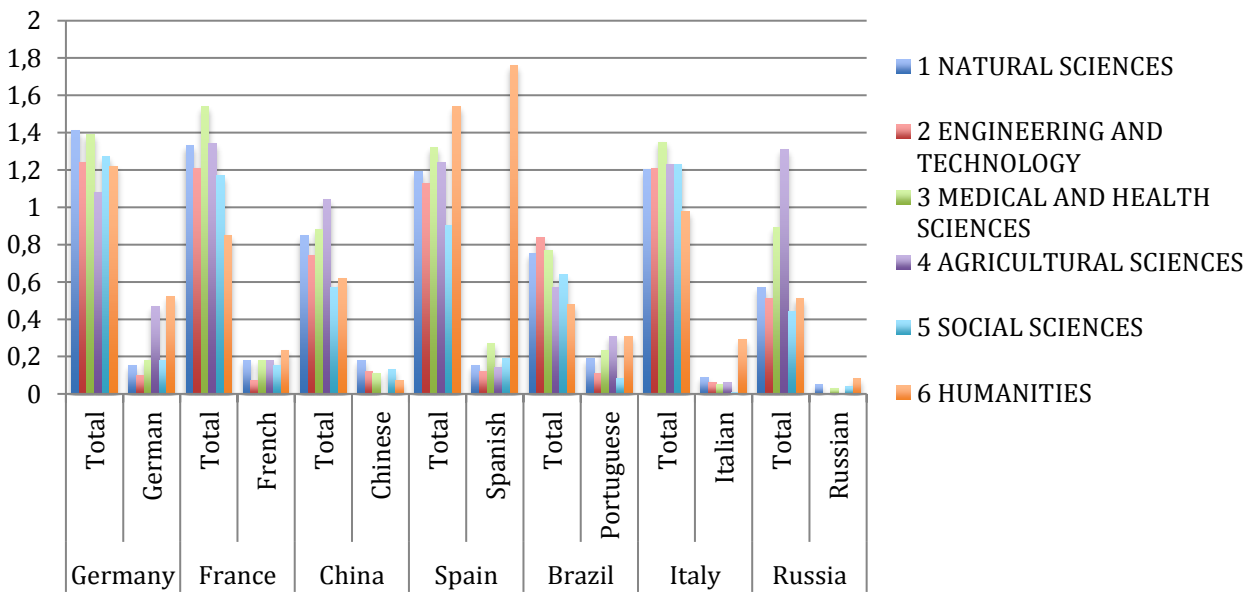

**Fig.10 Comparison of CNCI of publications in total and native-language publications (Web of Science, SCI-E, SSCI, AHCI, 2008-2016) for several countries**

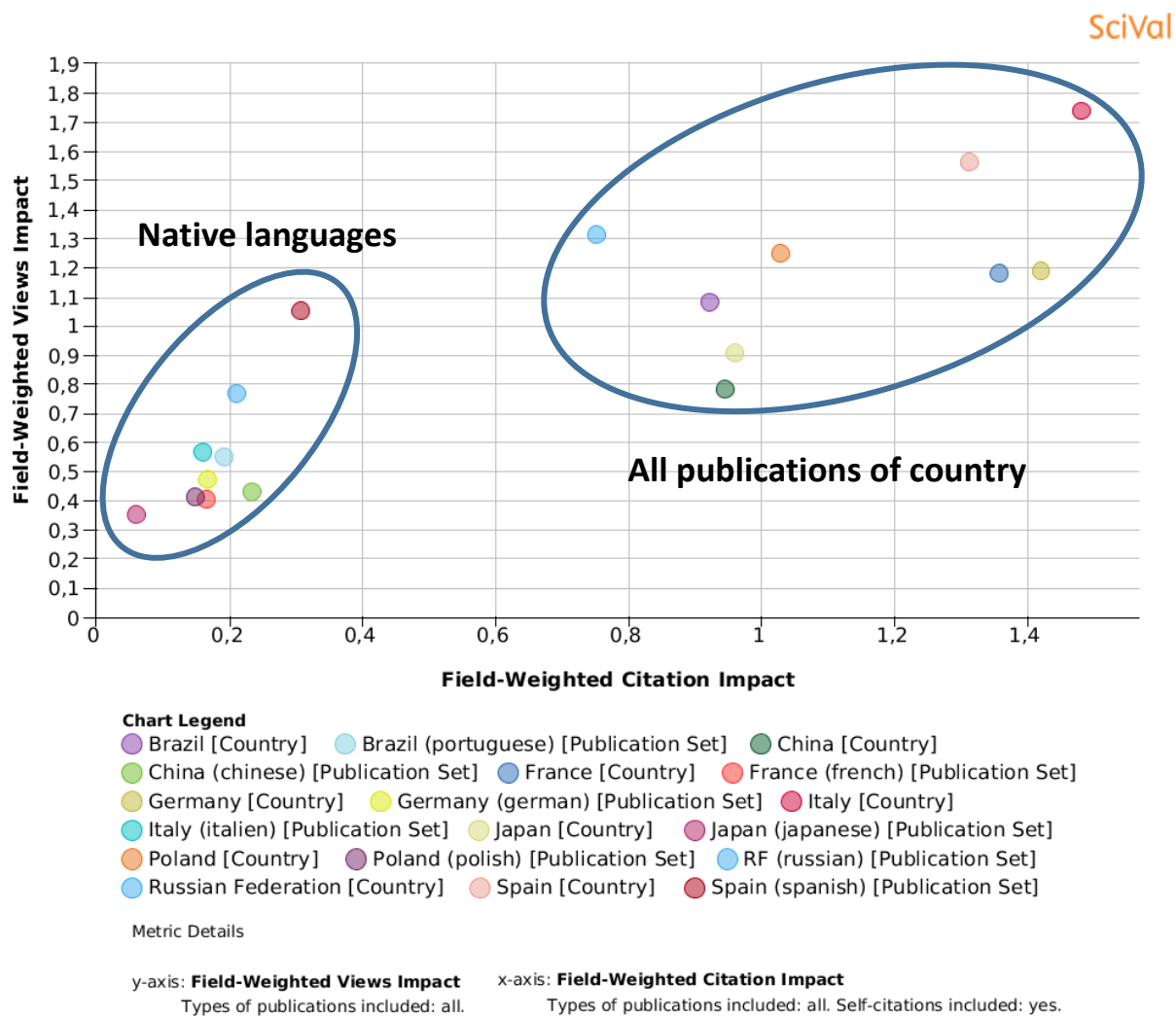

**Fig.11 Analysis of Field-Weighted Views Impact and Field-Weighted Citation Impact of countries' publications in total and publications in native language (Scopus, 2015-2016, SciVal data)**

The comparison of publications in native languages with all publications of certain countries in Scopus as analyzed in SciVal has demonstrated that Field-Weighted Views Impact and Field-Weighted Citation Impact are significantly lower for native-language publications (Fig.11). In each group we can see a correlation between Field-Weighted Views Impact and Field-Weighted Citation Impact, so we can propose that visibility of publications positively affect their citation. Citations are presumably from the countries where this language is the native one, i.e. citations of Russian-language publications of Russian Federation come from Russia itself and former Soviet Union countries – Belarus, Ukraine, Kazakhstan, etc. It is impossible to analyze the views by countries in SciVal, but we can suppose that their source is similar to that of citations.

**Ranking of Multilanguage journals**

Several publications demonstrate that English-language journals in total are ranked higher than non-English ones (Liang et al., 2013; Vinther & Rosenberg, 2012). For medical journals the dependence of IF and self-citation level on the number of English-language articles in Multilanguage journals (Diekhoff et al., 2013).

We tried to determine whether the journal ranking (journal IF) correlates with the share of English-language documents in journal for different research areas. We analyzed Multilanguage journals indexed in SCIE and SSCI of Web of Science CC. For the first step we have selected journals marked as Multilanguage from Master Journal List and chose those of them where the number of non-English language documents in 2006-2016 was more than 1. We have found 726 journals that met this criterion. The share of English-language documents in these journals is shown on Fig.12.

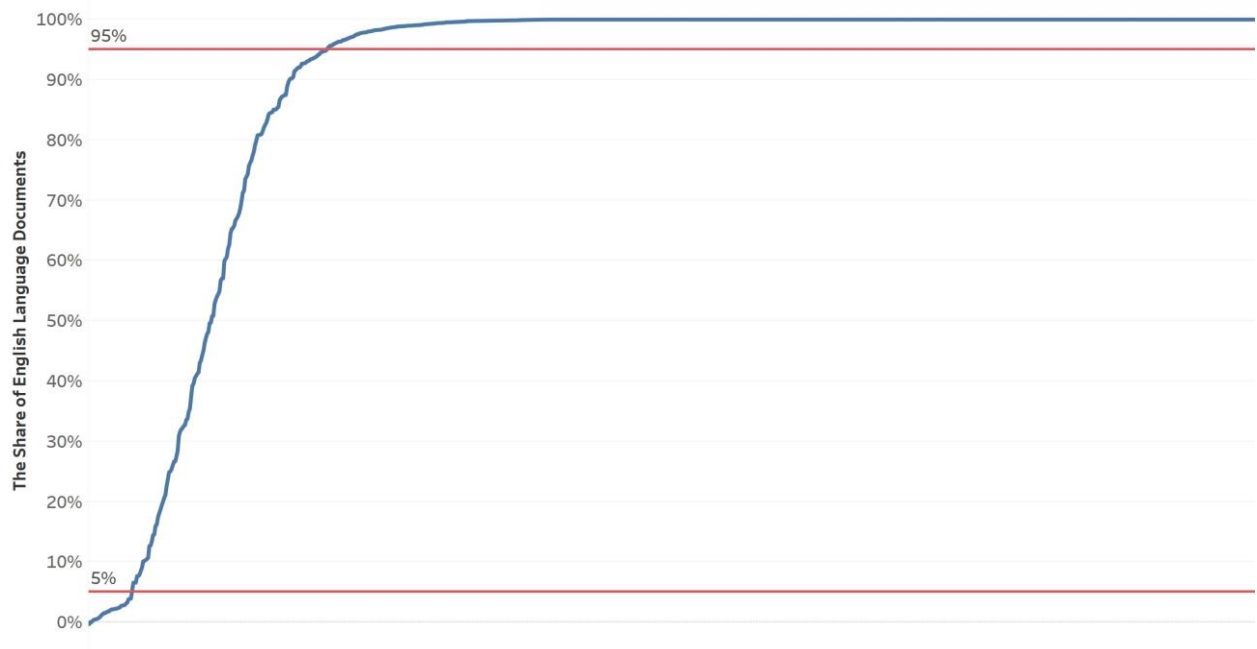

**Fig.12 Distribution of Multilanguage journals by proportion of English-language documents (Web of Science CC, 2006-2016).**

Later, 159 journals were selected from this list with the share of English-language documents from 5% to 95%. Annual number of English-language documents in 2006-2016 was calculated for each journal to compare these data with journal IF (IF-2017 is to be compared with share of English-language documents in 2015-2016). Selected journals belong to OECD research fields as demonstrated in Table 1.

We found that IF correlates positively with the share of English-language documents in journal not only in medical journals in (IF 2010 and share of English-language documents in 2008-2009) as it had been demonstrated in article mentions above (Diekhoff et al., 2013), but also for other fields of science. Fluctuations of the share of English-language documents during 10-year period influenced the values of IF of journals in natural sciences, engineering & technology and social sciences (Fig.13). Low correlation in humanities and absence of it in agriculture can be explained by peculiarities of citation in these areas and relatively local character of such journals. In agriculture, for example, almost a half of analyzed journals originates from Spanish and Portugal-speaking countries and about 60% of citations came from the same countries.

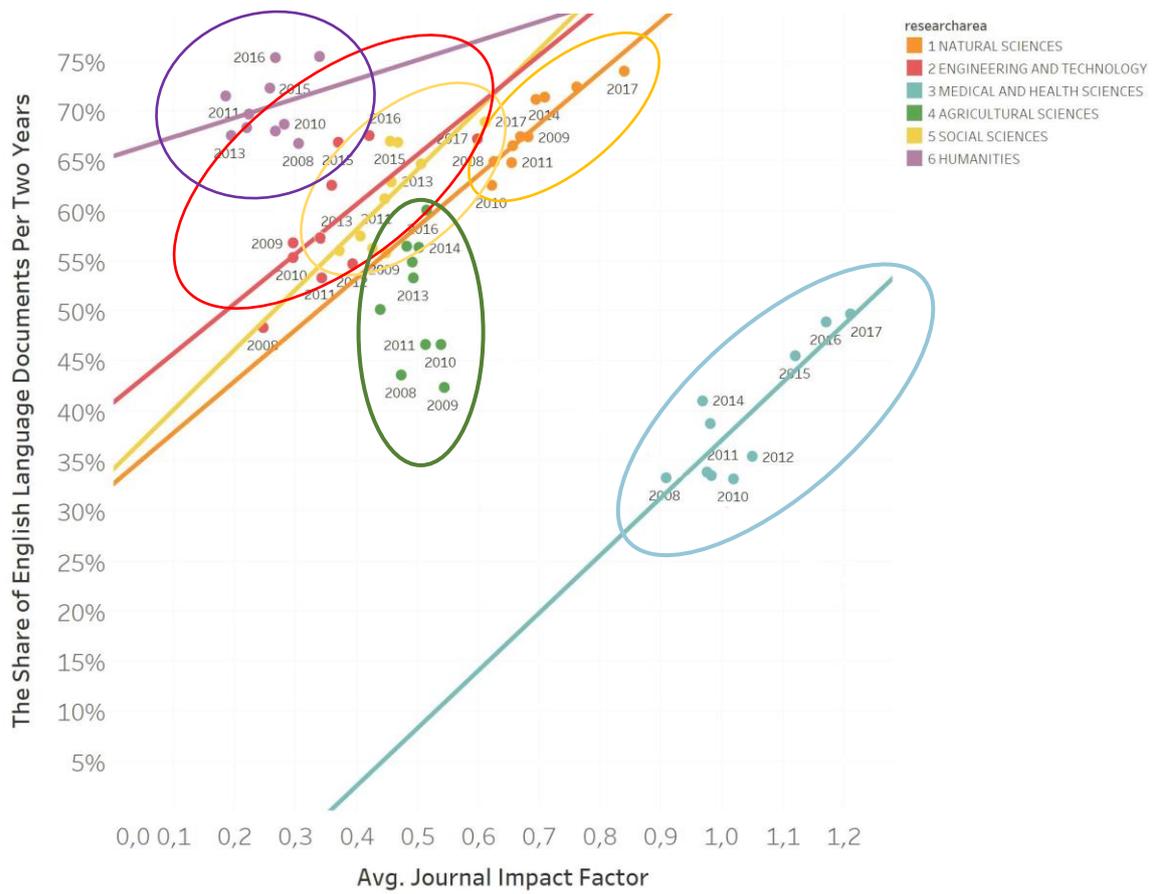

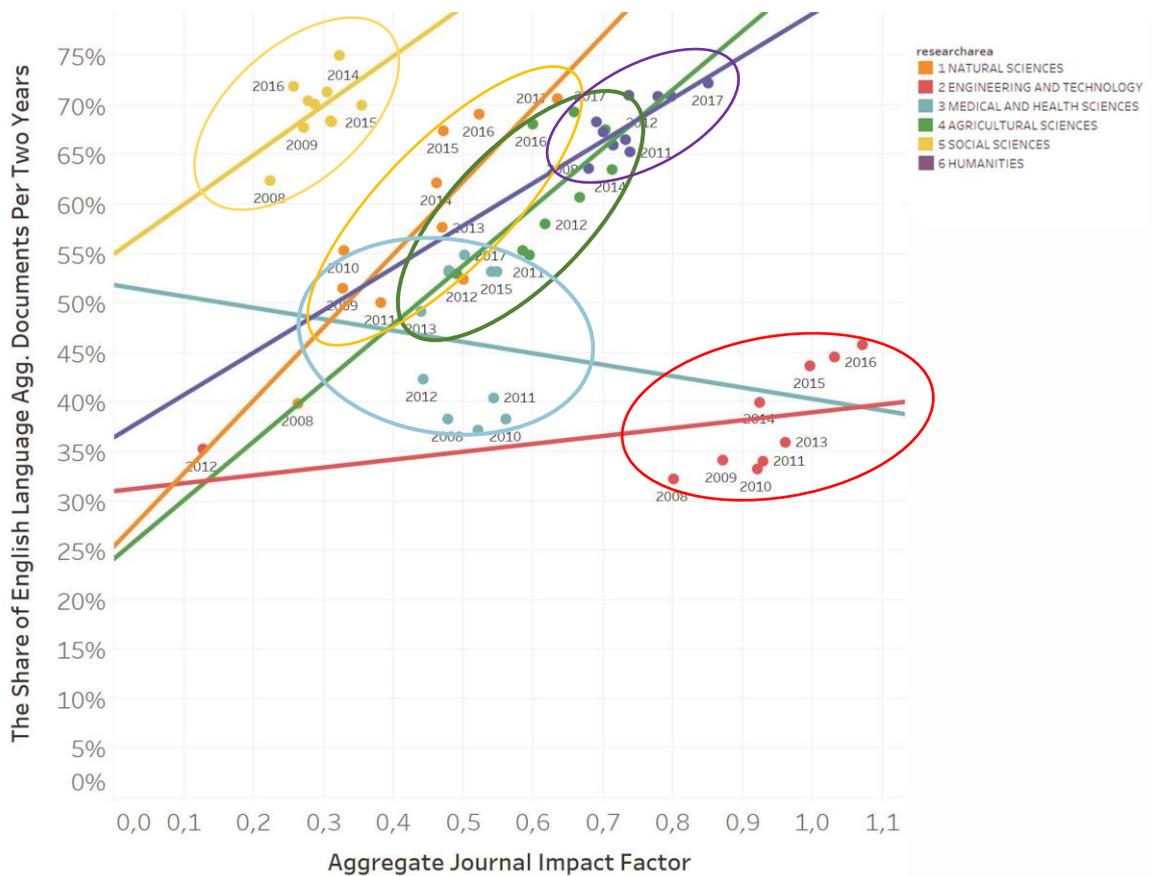

**Fig.13 Average (A) and aggregate (B) values of journal IF and share of English-language documents by OECD fields of science in 2006-2016.**

We understand that such analysis of IF and language dependence is rather crude, so we demonstrate comparison of average and aggregated impact factors for certain journal groups. The certain values for average and aggregate IF for analyzed journals are presented in tables 1(a, b)

**Table 1a Number of analyzed journals, average IF coefficients of correlation (Pearson) and P-values for OECD fields of science**

| OECD field of science | Journal number | $R^2$ | R | p-value |
|---|---|---|---|---|
| 1 Natural Sciences | 57 | 0.816 | 0,903 | 0,0003 |
| 2 Engineering and Technology | 10 | 0.577 | 0,760 | 0,0200 |
| 3 Medical Sciences | 31 | 0.773 | 0,879 | 0,0016 |
| 4 Agricultural Sciences | 22 | 0.042 | 0,205 | 0,5692 |
| 5 Social Sciences | 31 | 0.577 | 0,760 | 0,0108 |
| 6 Humanities | 8 | 0.089 | 0,298 | 0,4037 |

**Table 1b Number of analyzed journals, aggregate IF coefficients of correlation (Pearson) and P-values for OECD fields of science**

| OECD field of science | Journal number | R2 | R | p-value |
|---|---|---|---|---|
| 1 Natural Sciences | 57 | 0.6045 | 0.778 | 0.0081 |
| 2 Engineering and Technology | 10 | 0.6937 | 0,833 | 0.0027 |
| 3 Medical Sciences | 31 | 0.1754 | 0,419 | 0.2282 |
| 4 Agricultural Sciences | 22 | 0.0048 | 0,0693 | 0.8489 |
| 5 Social Sciences | 31 | 0.5447 | 0,738 | 0.01480 |
| 6 Humanities | 8 | 0.3097 | 0,556 | 0,095 |

Full list of analyzed journals, their distribution by OECD fields of science and annual details on IF and English-language share are available on Mendeley Data (Akoev, 2018).

**Conclusion**

Our results suggest the following:

- The role of English language in scientific publications increases evidently
- Native languages play essential role in social sciences and humanities
- Native-language publications are read and cited less than English-language outside the origin country
- The role of OA (Gold OA or Green OA) for non-English language publications (both views and citation) is even higher than average.
- The journal ranking for Multilanguage journals correlates with the share of English-language publications.

**Acknowledgments.**

**References**

Akoev, M. (2018). The list of Multi-Language Journals from Web of Science Core Collection masterlist

with Impact Factor and Share of English Language Documents In Previous Two YearsNo Title. https://doi.org/10.17632/r6dcf26mvp.2

Ammon, U. (2012). Linguistic inequality and its effects on participation in scientific discourse and on global knowledge accumulation – With a closer look at the problems of the second-rank language communities. *Applied Linguistics Review*, *3*(2), 333–355. https://doi.org/10.1515/applirev-2012-0016

Clarivate Analytics Partners with the Egyptian Knowledge Bank to Power the First Arabic Citation Index - Clarivate. (2018). Retrieved May 30, 2018, from https://clarivate.com/blog/news/clarivate-analytics-partners-egyptian-knowledge-bank-power-first-arabic-citation-index/?utm_source=linkedin&utm_medium=social&utm_campaign=EKB_ABM_ARCI_and_Professional_Services_SAR_EM_Egypt_2018

Di Bitetti, M. S., & Ferreras, J. A. (2017). Publish (in English) or perish: The effect on citation rate of using languages other than English in scientific publications. *Ambio*. https://doi.org/10.1007/s13280-016-0820-7

Diekhoff, T., Schlattmann, P., & Dewey, M. (2013). Impact of Article Language in Multi-Language Medical Journals - a Bibliometric Analysis of Self-Citations and Impact Factor. *PLoS ONE*, *8*(10), e76816. https://doi.org/10.1371/journal.pone.0076816

Ethnologue. (2015). Summary by language size: Language Size. *SIL International*. Retrieved from https://www.ethnologue.com/statistics/size

Fung, I. C. H. (2008). Citation of non-English peer review publications - Some Chinese examples. *Emerging Themes in Epidemiology*. https://doi.org/10.1186/1742-7622-5-12

González-Alcaide, G., Valderrama-Zurián, J. C., & Aleixandre-Benavent, R. (2012). The Impact Factor in non-English-speaking countries. *Scientometrics*, *92*(2), 297–311. https://doi.org/10.1007/s11192-012-0692-y

Gordin, M. D. (2015). *Scientific Babel. How Science Was Done Before and After Global English*. *The University of Chicago Press*.

Kulczycki, E., Engels, T. C. E., Pölönen, J., Bruun, K., Dušková, M., Guns, R., … Zuccala, A. (2018). Publication patterns in the social sciences and humanities: evidence from eight European countries. *Scientometrics*, 1–24. https://doi.org/10.1007/s11192-018-2711-0

Kumar, N., Panwar, Y., & Mahesh, G. (2016). A snapshot of research papers in non - English languages. *CURRENT SCIENCE*, *111*(1), 9–10. Retrieved from https://en.wikipedia.org/w/index.php?title=

Liang, L., Rousseau, R., & Zhong, Z. (2013). Non-English journals and papers in physics and chemistry: bias in citations? *Scientometrics*, *95*(1), 333–350. https://doi.org/10.1007/s11192-012-0828-0

Liu, F., Hu, G., Tang, L., & Liu, W. (2018). The penalty of containing more non-English articles. *Scientometrics*, *114*(1), 359–366. https://doi.org/10.1007/s11192-017-2577-6

Liu, W. (2017). The changing role of non-English papers in scholarly communication: Evidence from Web of Science's three journal citation indexes. *Learned Publishing*, *30*(2), 115–123. https://doi.org/10.1002/leap.1089

Meneghini, R., & Packer, A. L. (2007). Is there science beyond English? *EMBO Reports*, *8*(2), 112–116.

Montgomery, S. L. (2016). Impacts of a Global Language on Science: Are There Disadvantages? In *Language as a Scientific Tool. Shaping Scientific Language Across Time and National Traditions*. (pp. 199–218). Routledge. https://doi.org/10.4324/9781315657257

Moskaleva, O., & Akoev, M. (2018). Publications on different languages in citation indexes, or If there is a chance for Russian language in science? *Universitetskaya Kniga*, (3), 42–45.

Moskaleva, O., Pislyakov, V., Sterligov, I., Akoev, M., & Shabanova, S. (2018). Russian index of

Science citation: Overview and review. *Scientometrics*. https://doi.org/10.1007/s11192-018-2758-y

Pislyakov, V. (2007). What Are National Citation Indices Needed For ? *Scientific and Technical Libraries*, *2*, 66–68.

Shu, F., & Larivière, V. (2015). Chinese-language articles are biased in citations. *Journal of Informetrics*, *9*(3), 526–528. https://doi.org/10.1016/j.joi.2015.05.005

Sīle, L., Guns, R., Sivertsen, G., & Engels, T. C. E. (2017). European databases and repositories for Social Sciences and Humanities research output, (July). https://doi.org/10.6084/m9.figshare.5172322

UNESCO. (2016). *Unesco Science Report*. Retrieved from http://unesdoc.unesco.org/images/0023/002354/235406e.pdf

Vinther, S., & Rosenberg, J. (2012). Impact factor trends for general medical journals: non-English-language journals are lacking behind. *Swiss Medical Weekly 2012 142:3940*, *142*(3940). https://doi.org/10.4414/smw.2012.13572